\documentclass[useAMS,usenatbib]{mn2e}
\usepackage{txfonts}
\usepackage{pdflscape}
\usepackage{longtable}  
\usepackage{rotating}
\usepackage{natbib}
\usepackage{graphicx}
\usepackage{graphics}
\usepackage{psfrag}
\usepackage{amssymb}
\usepackage{multicol}
\bibpunct{(}{)}{;}{a}{}{,}
\def\Teff{\ensuremath{T_{\mathrm{eff}}}}
\def\logg{\ensuremath{\log g}}
\def\vmic{$\upsilon_{\mathrm{mic}}$}
\def\vmac{$\upsilon_{\mathrm{macro}}$}

\def\kms{$\mathrm{km\,s}^{-1}$}

\def\loggf{$\log{gf}$}
\def\vr{${\upsilon}_{\mathrm{r}}$}
\def\nlte{non-LTE}
\def\llm{{\sc LLmodels}}

\def\errTeff{$\sigma_{T_{\mathrm{eff}}}$}

\def\R{\ensuremath{R/R_{\odot}}}
\def\vald{{\sc VALD}}

\title[An In-Depth Spectroscopic Analysis of RR~Lyr]{An In-Depth Spectroscopic Analysis of RR~Lyr\\
  Variations over the Pulsation Cycle\thanks{Data obtained with the 2.7-m telescope at McDonald Observatory, Texas, US.} }
\author[L. Fossati, et al.]
{	L. Fossati$^{1}$\thanks{E-mail:lfossati@astro.uni-bonn.de},
	K. Kolenberg$^{2,3}$,
	D.~V. Shulyak$^{4}$,
	A. Elmasli$^{5}$,
	V. Tsymbal$^{6}$,
	T.~G. Barnes$^{7}$,
	\and 
	E. Guggenberger$^{8}$,
	O. Kochukhov$^{9}$
\\
\\
$^{1}$ Argelander-Institut f\"ur Astronomie der Universit\"at Bonn, Auf dem H\"ugel 71, 53121, Bonn, Germany\\
$^{2}$ Instituut voor Sterrenkunde, Celestijnenlaan 200D, B-3001 Heverlee, Belgium \\
$^{3}$ Harvard-Smithsonian Center for Astrophysics, 60 Garden Street, Cambridge MA 02138, USA \\
$^{4}$ Institute of Astrophysics, Georg-August-University, Friedrich-Hund-Platz 1, D-37077, G\"ottingen, Germany\\
$^{5}$ Department of Astronomy and Space Sciences, Ankara University, 06100, Tando\u{g}an, Ankara, Turkey\\
$^{6}$ Tavrian National University, Vernadskiy's Avenue 4, Simferopol, Crimea\\
$^{7}$ The University of Texas at Austin, McDonald Observatory, 2515 Speedway, Stop C1402, Austin, Texas 78712-1206, USA\\
$^{8}$ Institute of Astronomy, T\"urkenschanzstrasse 17, A-1180 Vienna, Austria\\ 
$^{9}$ Department of Physics and Astronomy, Uppsala University, BOX 516, SE-751 20, Uppsala, Sweden
}

\begin{document}

\date{}

\pagerange{\pageref{firstpage}--\pageref{lastpage}} \pubyear{2014}

\maketitle

\label{firstpage}

\begin{abstract}
The stellar parameters of RR~Lyrae stars vary considerably over a pulsation cycle, and their determination is crucial for stellar modelling. We present a detailed spectroscopic analysis of the pulsating star RR~Lyr, the prototype of its class, over a complete pulsation cycle, based on high-resolution spectra collected at the 2.7-m telescope of McDonald Observatory. We used simultaneous photometry to determine the accurate pulsation phase of each spectrum and determined the effective temperature, the shape of the depth-dependent microturbulent velocity, and the abundance of several elements, for each phase. The surface gravity was fixed to 2.4. Element abundances resulting from our analysis are stable over the pulsation cycle. However, a variation in ionisation equilibrium is observed around minimum radius. We attribute this mostly to a dynamical acceleration contributing to the surface gravity. Variable turbulent convection on time scales longer than the pulsation cycle has been proposed as a cause for the Blazhko effect. We test this hypothesis to some extent by using the derived variable depth-dependent microturbulent velocity profiles to estimate their effect on the stellar magnitude. These effects turn out to be wavelength-dependent and much smaller than the observed light variations over the Blazhko cycle: if variations in the turbulent motions are entirely responsible for the Blazhko effect, they must surpass the scales covered by the microturbulent velocity. This work demonstrates the possibility of a self-consistent spectroscopic analysis over an entire pulsation cycle using static atmosphere models, provided one takes into account certain features of a rapidly pulsating atmosphere.
\end{abstract}

\section{Introduction}
RR~Lyrae stars are used as standard candles and tracers of galactic evolution \citep[e.g.,][]{benedict2011}. They show mostly radial pulsations with large amplitudes, making them useful measures for theoretical modelling \citep{feu1999}. The modelling of pulsational signals requires the knowledge of stellar parameters, which can be derived from available observables (i.e., photometry and/or spectroscopy) using several methods, with varying degrees of reliability.

RR~Lyr, the prototype and eponym of its class, has been studied for over a century. As the nearest and brightest member of its class, it has been the subject of several spectroscopic studies, all of them analysing either a single spectrum obtained around minimum light, or a handful of spectra obtained at different Blazhko phases. The main goal of the present work is to perform a self-consistent analysis of RR~Lyr, deriving stellar parameters from a large number of high-resolution spectra, over a complete pulsation cycle. This is challenging in the phases of rapid atmospheric changes, where distorted spectral line profiles are observed, since the classical methods assume static atmospheres at each moment in time.

In a previous work \citep[][hereafter KF10]{KF10}, we concluded that the phase of maximum radius, when the spectral lines are least distorted, is the optimal phase for determining the abundances of the star in the most reliable way. We also developed a method to constrain the parameters of the star in the most robust way. In this paper we analyze all other phases in the pulsation cycle in a similar way as was described in KF10 and describe a single pulsation cycle of RR~Lyr on the basis of the quantities derived from the spectra.
\section{Observations}
A total of 64 spectra of RR~Lyr were obtained between June 26th and August 27th, 2004 with the Robert G. Tull Coud\'e Spectrograph (TS) attached to the 2.7-m telescope of McDonald observatory. This is a cross-dispersed \'echelle spectrograph yielding a resolving power of 60\,000 for the adopted configuration. The spectra cover the wavelength range 3633--10849\,\AA, with gaps between the orders at wavelengths greater than 5880\,\AA. 

Bias and flat-field frames were obtained at the beginning of each night, and a ThAr spectrum, for wavelength calibration, was often obtained during each night. The spectra were reduced using standard procedures with the Image Reduction and Analysis Facility \citep[IRAF\footnote{IRAF {\tt (http://iraf.noao.edu/)} is distributed by the National Optical Astronomy Observatory, which is operated by the Association of Universities for Research in Astronomy (AURA) under cooperative agreement with the National Science Foundation.},][]{tody}. Each spectrum was normalised by fitting a low order polynomial to carefully selected continuum points. The normalisation of the H$\gamma$ line was of crucial importance since we adopted the profile fitting of the H$\gamma$ line wings as a temperature indicator. We were unable to use either H$\alpha$ and H$\beta$ because the former was not covered by our spectra and the latter was affected by a spectrograph defect, preventing the normalisation. We were able to perform a reliable normalisation of the H$\gamma$ line using the artificial flat-fielding technique described by \citet{barklem02} and which we already successfully adopted with TS spectra \citep[see e.g., KF10;][]{fossati11,zwintz13}. 

An overview of the analysed McDonald spectra can be found in Table~\ref{tab:results}, which consists of a sub-sample of the data set presented in Table~1 of KF10, i.e. we selected only the spectra that are close in their Blazhko phase. The pulsation and Blazhko phases were calculated using simultaneous photometry presented by \citet{K06}. The pulsation and Blazhko phases, listed in Table~\ref{tab:results}, differ from those presented by KF10 where the phases were derived from ephemeris based on \citet{K06} data. The newly derived phases are more accurate, taking into account that the period of RR~Lyr varies over the Blazhko cycle \citep[see e.g., Fig.\,4 in][]{K11}. Integration times were generally 960\,s, hence less than 4\% of the pulsation cycle in order to avoid extreme phase smearing. On average our spectra have a S/N per pixel, calculated over 1\,\AA\ at $\sim$5000\,\AA, varying between 100 and 300.
\section{Data analysis}\label{sec:analysis}
We computed model atmospheres of RR~Lyr using the \llm\ stellar model atmosphere code \citep{llm}. For all calculations Local Thermodynamical Equilibrium (LTE) and plane-parallel geometry were assumed. We used the \vald\ database \citep{vald1,vald2,T83av} as a source of atomic line parameters for opacity calculations and abundance analysis. Convection was implemented according to the \citet{cm1,cm2} model of convection. More information regarding the adopted stellar model atmosphere are given in KF10.
\subsection{Previous analysis at the ``most stable pulsation phase''}
To better understand our analysis, we first summarise here the KF10 methodology and major findings. In KF10 we concluded that maximum radius is the optimal phase to perform a ``classical'' (i.e., assuming LTE, plane-parallel geometry and a static model atmosphere) spectroscopic analysis. We defined this as the ``quiet phase'' (QP). We therefore performed a detailed fundamental parameter determination and abundance analysis of the spectrum obtained closest to the QP (spectrum number 260), using both equivalent widths and synthetic spectra. We determined the effective temperature (\Teff) by fitting synthetic spectra to the observed H$\gamma$ line. The surface gravity (\logg) was derived by imposing the ionisation equilibrium for Fe \citep[taking into account an average 0.1\,dex \nlte\ correction for Fe\,{\sc i};][]{mashonkina}, Si, and Ti, and also making use of the wings of the Mg\,{\sc i}\,b lines. In imposing the equilibrium between the Fe\,{\sc i} line abundance and equivalent widths to determine the microturbulent velocity (\vmic), we noticed the need to introduce a depth dependent \vmic, which we derived assuming a third order polynomial. The considered metallic lines were measured mostly with equivalent widths, from which we derived the abundance of 45 elements (52 ions). For the QP we finally obtained \Teff\,=\,6125\,$\pm$\,50\,K, \logg\,=\,2.4\,$\pm$\,0.2, the depth-dependent \vmic\ profile (see Fig.~7 of KF10), and [Fe/H]\,=\,-1.30\,$\pm$\,0.10\,dex.
\subsection{Effective temperature}
\begin{figure*}
\begin{center}
\includegraphics[width=15cm,clip]{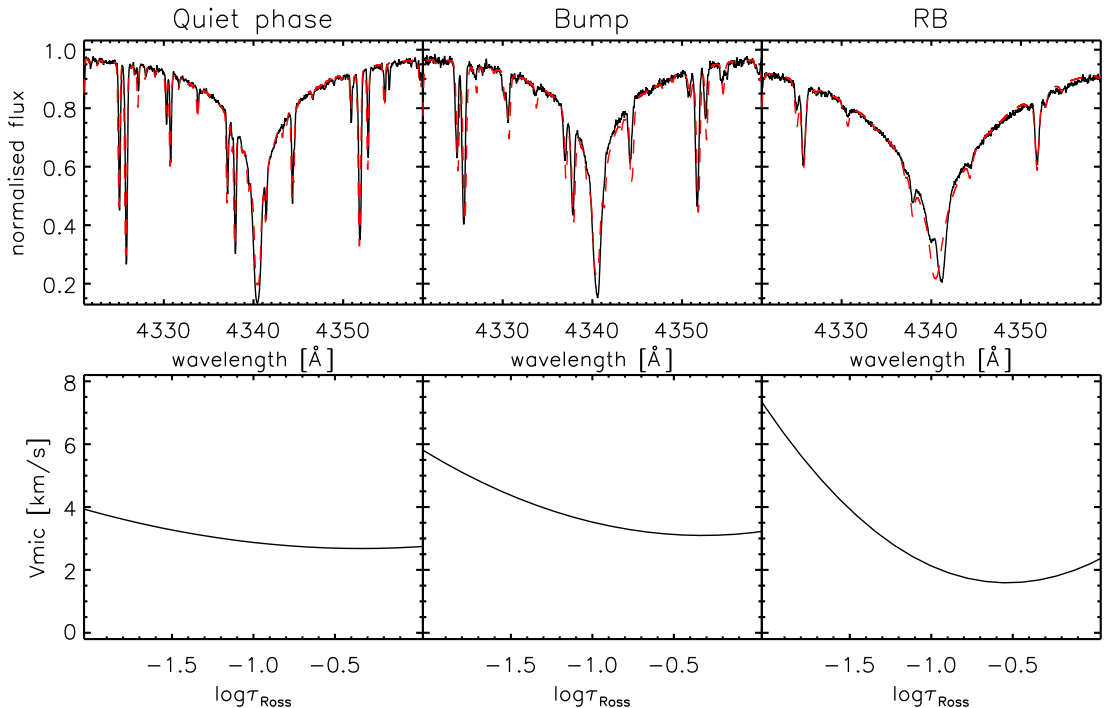}
\caption{Top panels: comparison between the observed H$\gamma$ line profile (black solid line) and a synthetic profile calculated with our final adopted parameters (red dashed line) for the spectra obtained at the quiet phase (QP; left), around the bump phase (centre), and on the rising branch (RB; right). Bottom panel: depth-dependent \vmic\ profile, around the main line-forming region, derived at the same three phases as in the top panels.}
\label{fig:hydrogen_vmic} 
\end{center} 
\end{figure*}

The effective temperature of RR~Lyr varies by almost 1000\,K during a pulsation cycle and as a consequence the \nlte\ effects vary with it: RR~Lyr is a metal-poor giant with a highly hydrodynamic atmosphere, hence strong non-LTE effects are expected to be present \citep[see e.g.,][]{mashonkina}. For this reason, there might be caveats in using the excitation equilibrium to compare \Teff\ values derived at different phases assuming LTE. We therefore determined \Teff\ by fitting synthetic spectra to the observed wings of the H$\gamma$ line, which are less subject to \nlte\ effects. During the pulsation cycle hydrogen lines modify their shape, but this affects just the line core, while the wings reflect only the variation in \Teff\  (\logg\ variations smaller than $\sim$0.5\,dex do not affect the hydrogen line wings in the temperature regime of RR~Lyr). The top panels of Fig.~\ref{fig:hydrogen_vmic} show a comparison between the observed H$\gamma$ line profile and a synthetic profile calculated with the final adopted atmospheric parameters for the quiet phase, around the bump phase \citep[a local maximum in the light curve associated with a shock wave shortly before minimum light,][]{Gillet98} and on the rising branch (shortly after minimum light, when the main shock occurs and the star reaches its minimum radius), respectively at phase 0.328 (spectrum number 260), 0.713 (spectrum number 161), and 0.951 (spectrum number 174). The spectrum obtained on the rising branch and shown in Fig.~\ref{fig:hydrogen_vmic} presents the largest distortions caused by the pulsation, as evidenced by the hydrogen line core splitting. Even in this extreme case, the line wings are well reproduced by the synthetic spectrum.
%
\subsection{Effective gravity}
As mentioned above, the strong pulsation distorts the line profiles along the pulsation cycle. Such distortions cannot be modelled yet and therefore it is not possible to use lines with developed wings to determine \logg. In addition, the varying \nlte\ effects do not allow one to safely adopt the ionisation equilibrium to derive \logg\ consistently for all phases. For this reason, we decided to fix \logg\ to 2.4, the value we obtained from the analysis of the QP (KF10). This assumption is also justified by the fact that the estimated radius variation of $\sim$0.5\,\R\  (see e.g., Fig.~1 of KF10) leads to a \logg\ variation of $\sim$0.1\,dex, smaller than the uncertainty on \logg. In addition, by adopting a constant \logg\ value one can obtain further information, on both \nlte\ effects and pulsation, by monitoring the ionisation equilibrium.
\subsection{Microturbulent velocity} 
For each analysed spectrum we also derived the depth-dependent \vmic\ profile, adopting the same procedure and tools described in KF10, except for the fact that we simultaneously used all measured lines of Mg\,{\sc i}, Ca\,{\sc i}, Ti\,{\sc ii}, Cr\,{\sc i}, Fe\,{\sc i}, Fe\,{\sc ii}, and Ba\,{\sc ii}, while only Fe\,{\sc i} lines were used in KF10. This allowed us to determine a \vmic\ profile which best reproduce the observations at all phases. The symmetric line profiles of the spectrum taken at the QP allowed us to measure the equivalent widths assuming Gaussian line profiles and therefore to solve the simple blends. On the other hand, the other available spectra show distorted line profiles, which cannot be reproduced with an analytical line profile shape, hence we measured only the completely unblended lines by direct integration. The use of several ions in the determination of the depth-dependent \vmic\ allowed us to obtain a more robust solution, also because they provide a larger spread in equivalent width compared to what is available by using Fe\,{\sc i} only. The \vmic\ profiles around the main line-forming region derived at the QP, bump phase, and rising branch are shown in the bottom panels of Fig.~\ref{fig:hydrogen_vmic}. As expected, the turbulence is minimum at the QP and maximum on the rising branch, when the shock wave passes through the atmosphere.

It is important to remark that we used a third order polynomial to derive the depth-dependent \vmic\ profile. The polynomial fit is constrained only in the region of the atmosphere where metallic lines are formed (i.e., $-4\,\lesssim\,\log\tau_{Ross}\,\lesssim\,0$), while outside this region the \vmic\ profile is unreliable as it depends upon the polynomial extrapolation.
\subsection{Abundances}
To further check our results, we derived an independent set of abundances for each analysed spectrum. For all measured lines we used the same set of atomic parameters as in KF10. For consistency, we derived the atmospheric parameters (\Teff, \logg, \vmic, and abundances) iteratively, taking all of them into account, even in the computation of the stellar atmosphere models. This is particularly important as KF10 showed that a depth-dependent \vmic\ has a non-negligible impact on the atmospheric structure.
\subsection{Comparison with the KF10 results} 
\begin{figure}
\begin{center}
\includegraphics[width=85mm,clip]{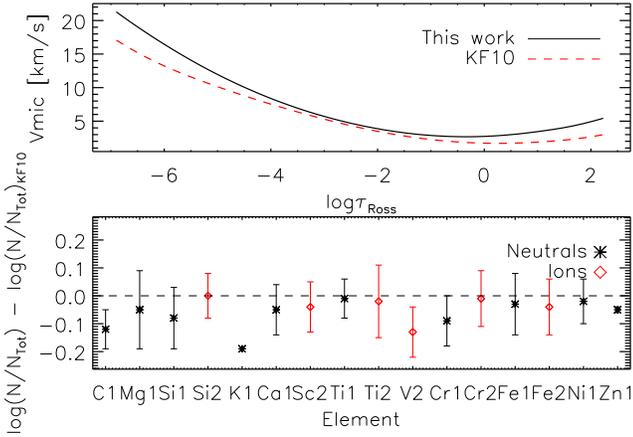}
\caption{Top panel: comparison between the \vmic\ profiles derived in this work (black solid line) and in KF10 (red dashed line) for the QP. Bottom panel: comparison between the abundances obtained in this work and by KF10 at the QP. The neutral elements are displayed as black asterisks, while ions are displayed as red rhombs. The uncertainties are the KF10 internal scatter uncertainties.}
\label{fig:vmic_comparison} 
\end{center} 
\end{figure}

Since we adopted a slightly different methodology than in KF10, we compare here the results we obtained for the QP with that of KF10. For the QP we derived \Teff\,=\,6100\,$\pm$\,50\,K and [Fe/H]\,=\,-1.34\,$\pm$\,0.09\,dex, in good agreement with that given in KF10. As the major difference is in the \vmic\ determination, in the top panel of Fig.~\ref{fig:vmic_comparison} we compare the \vmic\ profile we obtained for the QP to that of KF10. The average difference between the two profiles is $\sim$1.5\,\kms, but it reduces to $\sim$0.7\,\kms\ in the main line-forming region between $\log\tau_{Ross} -4$ and $0$. The difference between the two curves is most likely due to the different set of adopted lines; in particular, the use of some Ti\,{\sc ii} and Ba\,{\sc ii} lines which are stronger than the strongest Fe\,{\sc i} lines measured by KF10. As shown in KF10, the deviation from the ``constant \vmic\ equilibrium'' increases with increasing line strength (equivalent width) and the use of very strong lines provides further constraints on the outermost layers, not sampled by the set of lines adopted by KF10. In addition, the simultaneous use of various ions, rather than Fe\,{\sc i} only, allows one to obtain a more consistent description of the general atmospheric properties.

The bottom panel of Fig.~\ref{fig:vmic_comparison} shows the comparison between the abundances obtained in this work and by KF10. Beyond the general good agreement, one can notice a slight systematic shift. This is due to the higher/steeper \vmic\ profile; in particular, the potassium abundance is derived from only one rather strong line and therefore its abundance value is very dependent upon the adopted \vmic\ profile.
%
\section{Results }
\subsection{Effective temperature}
Table~\ref{tab:results} lists, for each analysed spectrum, the \Teff\ value derived from the H$\gamma$ line profile fitting. The uncertainty is mostly based on the S/N of the spectrum and on the uncertainty on the normalisation, while it neglects the model uncertainties (e.g., plane-parallel assumption). Table~\ref{tab:results} shows also that on average, as \Teff\ increases so does its uncertainty. This is because, with increasing \Teff, the hydrogen lines become less sensitive to \Teff\ variations. Except for the phases around maximum light, where the Blazhko modulation shows its maximum effect, the agreement between \Teff\ values derived from spectra obtained on different nights, but at similar pulsation phases increases the confidence in our results. 
\begin{table}
\caption[ ]{Basic data of the analysed observations of RR~Lyr in order of pulsation phase. The first column shows the pulsation phase and the second column the spectrum ID number. The third column lists the Heliocentric Julian Date (HJD-2453000) at the beginning of the exposure. The fourth and fifth columns list the Blazhko phase and the S/N per pixel, calculated at $\sim$5000\,\AA. The seventh column gives the exposure time in seconds. The last two columns list the \Teff\ value and its uncertainty obtained from the fitting of the H$\gamma$ line. The spectra number 253 and 257 have a lower exposure time, because the observation was stopped due to the presence of thick clouds.}
\label{tab:results}
\begin{center}
\scriptsize{
\begin{tabular}{lccccccc}
\hline
\hline
Pulsation & Sp. ID & HJD $-$ & Blazhko & S/N per  & Exposure & \Teff & \errTeff \\	
phase     & number & 2453000 & phase   & pixel    & time [s] &  [K]  &  [K]     \\
\hline
0.009 & 126 & 184.8121 & 0.306 & 63  & 960.000 & 7050 & 100 \\
0.023 & 178 & 185.9512 & 0.336 & 357 & 960.000 & 7050 & 75  \\
0.148 & 251 & 187.7208 & 0.381 & 359 & 960.000 & 6525 & 50  \\
0.170 & 252 & 187.7330 & 0.382 & 319 & 960.000 & 6450 & 50  \\
0.189 & 253 & 187.7442 & 0.382 & 120 & 769.397 & 6400 & 75  \\
0.213 & 087 & 183.7953 & 0.280 & 293 & 960.000 & 6325 & 50  \\
0.223 & 255 & 187.7632 & 0.382 & 124 & 960.000 & 6325 & 75  \\
0.247 & 088 & 183.8147 & 0.281 & 250 & 960.000 & 6275 & 50  \\
0.253 & 256 & 187.7803 & 0.383 & 160 & 960.000 & 6250 & 50  \\
0.269 & 089 & 183.8271 & 0.281 & 172 & 960.000 & 6225 & 50  \\
0.274 & 257 & 187.7923 & 0.383 & 87  & 797.949 & 6200 & 75  \\
0.299 & 258 & 187.8060 & 0.383 & 272 & 960.000 & 6175 & 50  \\
0.299 & 091 & 183.8441 & 0.281 & 149 & 960.000 & 6175 & 100 \\
0.328 & 260 & 187.8226 & 0.384 & 322 & 960.000 & 6100 & 50  \\
0.353 & 261 & 187.8370 & 0.384 & 214 & 960.000 & 6050 & 75  \\
0.377 & 262 & 187.8504 & 0.385 & 288 & 960.000 & 6025 & 50  \\
0.396 & 204 & 186.7293 & 0.356 & 190 & 960.000 & 6050 & 50  \\
0.400 & 263 & 187.8634 & 0.385 & 216 & 960.000 & 6025 & 50  \\
0.420 & 205 & 186.7425 & 0.356 & 138 & 960.000 & 6000 & 50  \\
0.441 & 206 & 186.7548 & 0.356 & 203 & 960.000 & 5950 & 75  \\
0.463 & 207 & 186.7670 & 0.357 & 46  & 960.000 & 5950 & 150 \\
0.499 & 209 & 186.7876 & 0.357 & 131 & 960.000 & 5975 & 75  \\
0.522 & 210 & 186.8007 & 0.358 & 77  & 960.000 & 5950 & 75  \\
0.649 & 158 & 185.7401 & 0.330 & 101 & 960.000 & 6000 & 50  \\
0.670 & 159 & 185.7523 & 0.331 & 126 & 960.000 & 6025 & 50  \\
0.692 & 160 & 185.7645 & 0.331 & 171 & 960.000 & 6050 & 50  \\
0.713 & 161 & 185.7768 & 0.331 & 233 & 960.000 & 6050 & 50  \\
0.743 & 163 & 185.7937 & 0.332 & 232 & 960.000 & 6025 & 50  \\
0.765 & 164 & 185.8059 & 0.332 & 259 & 960.000 & 6050 & 50  \\
0.787 & 165 & 185.8182 & 0.332 & 113 & 960.000 & 6025 & 50  \\
0.808 & 166 & 185.8304 & 0.333 & 210 & 960.000 & 6000 & 50  \\
0.837 & 168 & 185.8465 & 0.333 & 228 & 960.000 & 6025 & 50  \\
0.858 & 169 & 185.8587 & 0.333 & 290 & 960.000 & 6025 & 50  \\
0.863 & 119 & 184.7293 & 0.304 & 215 & 960.000 & 6025 & 100 \\
0.880 & 170 & 185.8709 & 0.334 & 236 & 960.000 & 6050 & 50  \\
0.888 & 120 & 184.7434 & 0.305 & 253 & 960.000 & 6125 & 100 \\
0.901 & 171 & 185.8831 & 0.334 & 216 & 960.000 & 6275 & 75  \\
0.910 & 121 & 184.7559 & 0.305 & 271 & 960.000 & 6375 & 100 \\
0.929 & 173 & 185.8989 & 0.334 & 236 & 960.000 & 6575 & 75  \\
0.932 & 122 & 184.7683 & 0.305 & 194 & 960.000 & 6725 & 100 \\
0.951 & 174 & 185.9112 & 0.335 & 215 & 960.000 & 6925 & 100 \\
0.964 & 124 & 184.7865 & 0.306 & 162 & 960.000 & 7050 & 100 \\
0.972 & 175 & 185.9234 & 0.335 & 233 & 960.000 & 7125 & 75  \\
0.986 & 125 & 184.7988 & 0.306 & 126 & 960.000 & 7125 & 100 \\
0.994 & 176 & 185.9356 & 0.335 & 290 & 960.000 & 7125 & 75  \\
\hline
\end{tabular}}
\end{center}
\end{table}

%
\subsection{Microturbulent velocity}
\begin{figure}
\begin{center}
\includegraphics[width=85mm,clip]{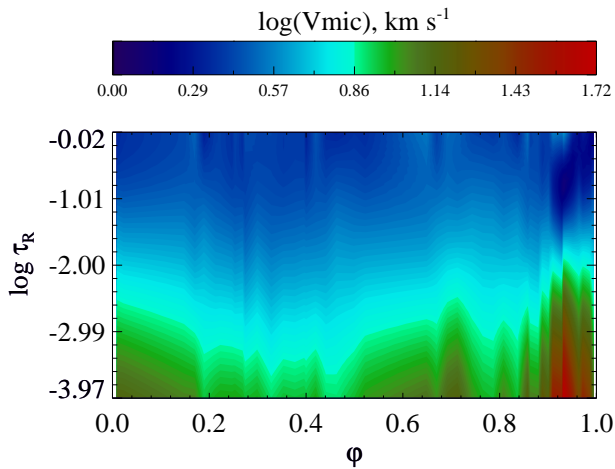}
\caption{Two dimensional plot of the depth-dependent \vmic\ profile (in log scale), obtained in the main line forming region, as a function of phase $\phi$.}
\label{fig:vmic_2D} 
\end{center} 
\end{figure}

From each analysed spectrum we also derived the depth-dependent \vmic\ profile. The bottom panels of Fig.~\ref{fig:hydrogen_vmic} show the \vmic\ profile obtained at three markedly different points in the pulsation cycle. The bump and rising branch are known to be connected with an increase in turbulence, as is also reflected in Fig.~4 of KF10, showing local extrema in the full width at half-maximum (FWHM) of different spectral lines in RR~Lyr's spectrum. Such increase in turbulence is clearly visible in Fig.~\ref{fig:hydrogen_vmic}, when comparing the \vmic\ profiles obtained at the QP with the other two. Figure~\ref{fig:vmic_2D} shows a 2D plot of all derived depth-dependent \vmic\ profiles, obtained in the main line-forming region, as a function of phase. We will further discuss the derived \vmic\ profiles in Sec.~\ref{vmic}. 
%
\subsection{Abundances}
Table~\ref{tab:abn} lists the Mg\,{\sc i}, Si\,{\sc i}, Si\,{\sc ii}, Ca\,{\sc i}, Ti\,{\sc i}, Ti\,{\sc ii}, Cr\,{\sc i}, Cr\,{\sc ii}, Fe\,{\sc i}, Fe\,{\sc ii}, Ni\,{\sc i}, and Ba\,{\sc ii} abundances obtained from the analysis of each analysed spectrum. As expected, the abundances remain about constant, within the uncertainties, throughout the whole pulsation cycle, with a few exceptions which will be discussed in detail in Sect.~\ref{sec:abn}.
\section{Discussion}
\subsection{Effective temperature and radius variation} 
Fundamental parameters of RR~Lyr at different pulsation phases were obtained by several authors using different methods \citep[e.g.,][]{siegel82,lambert96,takeda06}. The published fundamental parameters of RR~Lyr display a considerable scatter both in \Teff\ and \logg, mostly due to the large pulsation amplitudes and to the use of different assumptions (e.g., model atmosphere codes and photometric calibrations). According to these analyses, RR~Lyr's \Teff\ varies over its 13\,h~36\,min pulsation cycle between approximately 6250 and 8000\,K and its \logg\ between 2.4 and 3.8. 

Moreover, the Blazhko modulation leads to an additional strong variation of the fundamental parameters. \citet{jurcsik09} showed that also the {\it mean} properties of modulated RR~Lyrae stars change over the Blazhko cycle. \citet{sodor09} developed a new method for determining physical parameters of fundamental-mode RR~Lyrae stars from multicolour light curves, called the Inverse Photometric Method (IPM). A spectroscopic analysis performed along the complete pulsation cycle, as that presented in this work, will be the only way to test and calibrate the IPM. A future spectroscopic test and confirmation of the IPM would be most valuable as multicolour photometric data of RR~Lyrae stars are much more readily available than spectra. Moreover, it would confirm the observation by \citet{jurcsik09} beyond doubt, and hence facilitate more quantifiable testing of the models for the Blazhko effect.
\begin{figure*}
\includegraphics[width=17cm,clip]{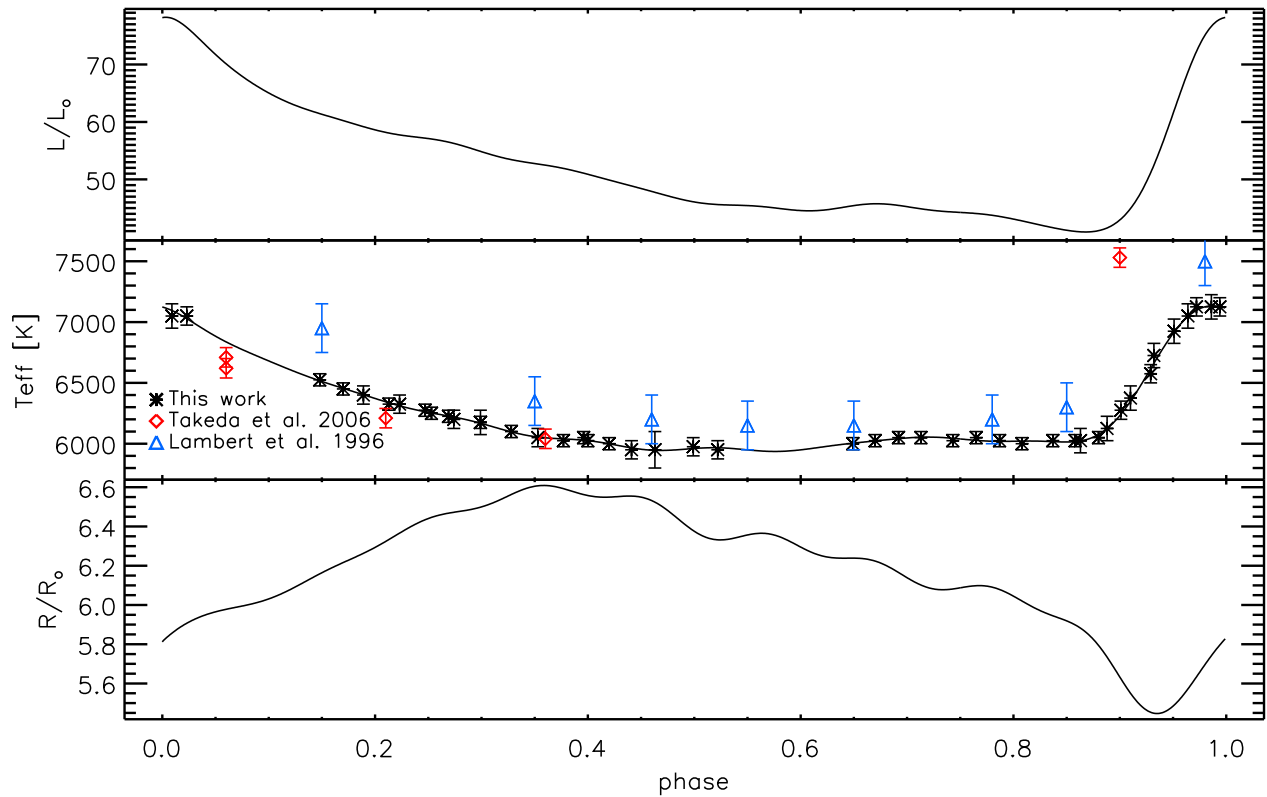}
\caption{Variation of luminosity (upper), \Teff\  (middle), and radius (bottom) over the RR~Lyr pulsation cycle. The solid line in the middle panel is a fit to the derived \Teff\ value. The blue triangles and the red diamonds show the \Teff\ values obtained by \citet{lambert96} and \citet{takeda06}, respectively.}
\label{fig:LTR}
\end{figure*}

Figure~\ref{fig:LTR} shows the luminosity, \Teff, and stellar radius variation over the pulsation cycle. We derived the luminosity from the Johnson $V$-band magnitude \citep[from][]{K06}, correcting it for a distance of 262\,pc \citep{benedict02}, a reddening of A$_V$=0.007, and by applying the bolometric correction by \citet{balona}. The \Teff\ variation over the whole pulsation cycle was determined by interpolating the measured \Teff\ values, while the stellar radius was then derived from the luminosity and \Teff. It is important to remark here that the radius determined in this way is that of the layer at which the effective temperature is defined \citep[see e.g.,][for an extreme case]{langer1989}. The wiggling of the stellar radius, shown in Fig.~\ref{fig:LTR} at phases between 0.35 and 0.8 is most likely an artifact of the scatter in the \Teff\ values. 

The \Teff\ uncertainties obtained from the H$\gamma$ line profile fitting should be considered on a relative scale (comparison of one \Teff\ value to the neighbouring ones), while on an absolute scale the uncertainties should be taken with caution, because of the model assumptions; in particular, the absolute \Teff\ uncertainties should be larger around maximum light, where the main shock wave is passing through the atmosphere. Nevertheless, previous modelling \citep{fokin92} and tests we performed with synthetic spectra and simulated temperature profiles accounting for the shock wave suggest that the wings of hydrogen lines are barely affected by hydrodynamic effects, improving our confidence in the derived \Teff\ values, also on an absolute scale. 

We obtained a \Teff\ variation in phase with the luminosity (i.e., maximum light and maximum temperature occur simultaneously). Our results show also that the phase of minimum radius occurs shortly before maximum light. Both these findings are in agreement with hydrodynamic modelling of the RR~Lyr pulsation cycle (e.g., see Fig~1 of KF10). 

By calibrating Stroemgren photometry, \citet{siegel82} derived the \Teff\ variation along almost the whole RR~Lyr pulsation cycle. The shape of this variation is similar to the one we obtained, while their values span between a minimum of $\sim$6400\,K and a maximum of $\sim$8000\,K. Both values are higher  and span a far larger \Teff\ range, compared to our results. This large difference may be partly due to the improvement of stellar model atmospheres (on which the photometric calibrations are based), in particular the opacities. In addition, cycles at different Blazhko phases will likely yield measurably different temperature variations \citep[see also][]{jurcsik09}. 

Figure~\ref{fig:LTR} shows a comparison between the \Teff\ values we obtained from the H$\gamma$ line profile fitting with that published by \citet{lambert96} and \citet{takeda06}. One can notice a satisfactory good agreement between the three \Teff\ determinations, despite the fact that the Blazhko modulation spreads the measurements obtained at different Blazhko phases, particularly those close to maximum light, where the Blazhko effect has a stronger impact. In this comparison it is important to notice that \citet{takeda06} derived \Teff\ spectroscopically (from the excitation equilibrium of Fe\,{\sc i} lines and H$\alpha$ line profile fitting), while \citet{lambert96} adopted the average value derived from several photometric temperature calibrations, including spectroscopy (i.e., Fe\,{\sc i} excitation equilibrium).

\citet{for11} derived the effective temperature at different pulsation phases for a sample of RR~Lyrae stars using the Fe\,{\sc i} and Fe\,{\sc ii} excitation equilibrium. Their sample comprises both Blazhko and non-Blazhko stars and it did not contain RR~Lyr itself. For all stars they obtained an amplitude and shape of the variation similar to that shown in Fig.~\ref{fig:LTR}.
\subsection{Abundances}\label{sec:abn}
\begin{figure*}
\includegraphics[width=17cm,clip]{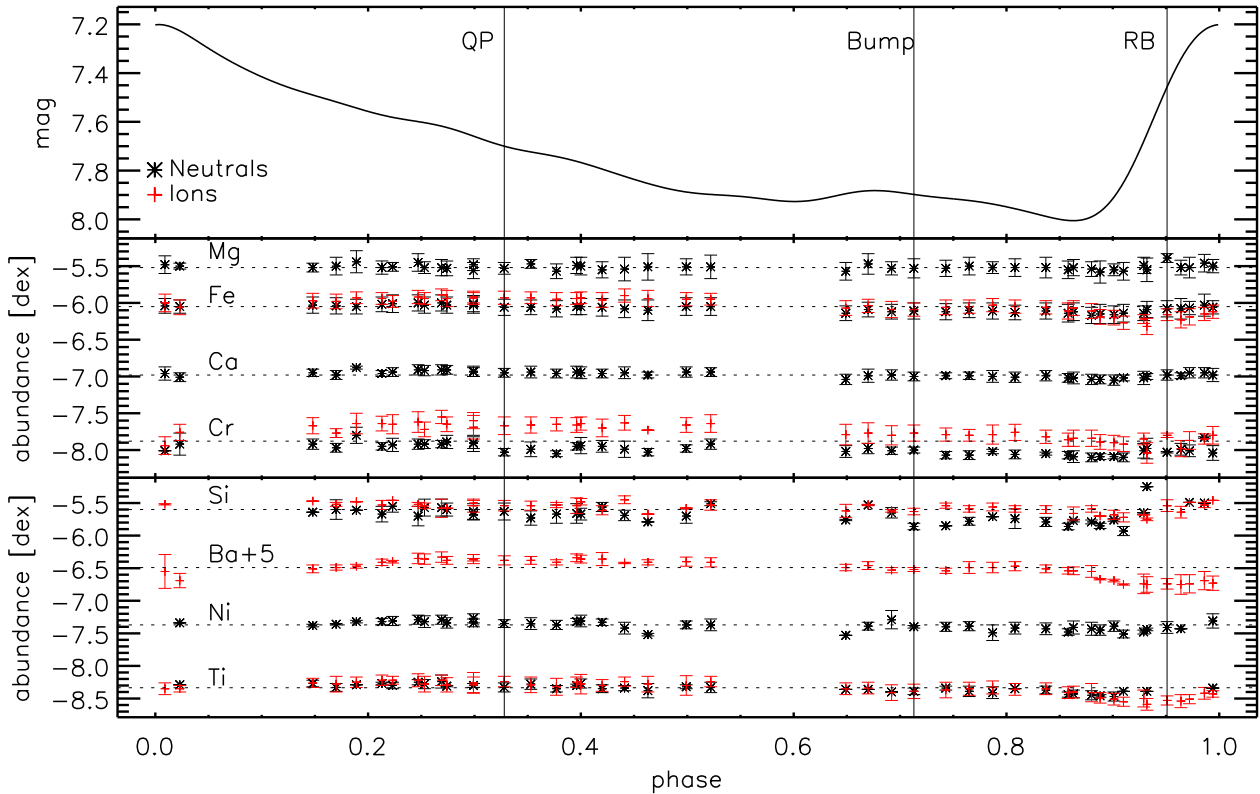}
\caption{Element abundances for Mg\,{\sc i}, Fe\,{\sc i}, Fe\,{\sc ii}, Ca\,{\sc i}, Cr\,{\sc i}, Cr\,{\sc ii} (middle panel), Si\,{\sc i}, Si\,{\sc ii}, Ba\,{\sc ii}, Ni\,{\sc i}, Ti\,{\sc i}, and Ti\,{\sc ii} (bottom panel) derived over the pulsation cycle. The abundances obtained for Ba\,{\sc ii} were shifted upwards by 5\,dex for visualisation reasons. The abundances of the neutral elements are indicated by black asterisks, while the abundances of the singly ionised elements are shown by red pluses. The dashed horizontal lines show the median of the abundances obtained over the pulsation cycle for each element. The upper panel shows the variation of the Johnson $V$-band magnitude obtained from \citet{K06}. The vertical solid lines indicate the position of the quiet phase (maximum radius), the phase of the bump, and a phase on the rising branch, as described in Sect.~\ref{sec:analysis}. The rather large difference between the Cr\,{\sc i} and Cr\,{\sc ii} abundance is mostly due to the choice of the Cr\,{\sc ii} \loggf\ values \citep{fossati11}.}
\label{fig:abn}
\end{figure*}

\citet{takeda06} were the first to derive the abundance of several elements as a function of pulsation phase for RR~Lyr, obtaining, as expected, that the  abundances remain constant along the pulsation cycle. Figure~\ref{fig:abn} shows our derived Mg\,{\sc i}, Fe\,{\sc i}, Fe\,{\sc ii}, Ca\,{\sc i}, Cr\,{\sc i}, Cr\,{\sc ii}, Si\,{\sc i}, Si\,{\sc ii}, Ba\,{\sc ii}, Ni\,{\sc i}, Ti\,{\sc i}, and Ti\,{\sc ii} abundance as a function of pulsation phase. Except for the region around the rising branch, the abundance of the analysed elements remains constant along the pulsation cycle, as one would expect. This result confirms that on a relative scale the derived atmospheric parameters are free from systematics. In addition, as our abundances have been obtained by assuming a constant \logg\ value of 2.4, the fact that there are no major variations in the ionisation equilibrium along the pulsation cycle let us conclude that the amplitude of the \logg\ variation is of the order of 0.1\,dex or smaller. This agrees with pulsation models (see KF10) and validates our assumption of a constant \logg\ value. Note that this assumption is also validated by the results obtained by \citet{for11} who spectroscopically analysed a sample of RR\,Lyrae pulsators keeping \logg\ as a free parameter. For the majority of the stars they analysed the maximum variation between the abundance of two ions of the same element was within the uncertainties.

Figure~\ref{fig:abn} shows that the abundance of the singly ionised elements decreases after minimum light, reaches a minimum in the middle of the rising branch and goes back to the average value at maximum light. We attribute this phenomenon to a variation of the dynamical acceleration term in the effective gravity \citep[see][for a definition of effective gravity]{for11}. As a matter of fact, during the rising branch the shock wave propagates rapidly through the atmosphere leading to a non-equilibrium condition driven by strong dynamics in the atmosphere of the star. From our results we obtain that the maximum deviation from the Fe ionisation equilibrium (taking into account \nlte\ effects) is reached by increasing \logg\ by 0.3\,dex, in agreement with the results obtained by \citet{for11} for other RR\,Lyrae pulsators.
\subsection{Microturbulent velocity}\label{vmic}
Turbulent velocity variations, that we measured from line equivalent widths throughout the pulsation cycles of RR\,Lyr, are reflected also in the variation of the spectral lines' full width at half maximum (FWHM), as can be clearly seen in Fig.~4 of KF10. As a consequence, the microturbulent velocity, which empirically reflects motions on scales smaller than the line-forming region, shows a behaviour similar to that of the FWHM. Usually one parametrises these motions with a constant \vmic\ value throughout the entire atmosphere, while it is clear that this parameter can vary with depth \citep[][KF10]{takeda06}. In our previous analysis of the spectrum obtained at the phase of maximum radius, we found that the use of a depth-dependent \vmic\ leads to a considerable improvement in the fit of several metallic lines, particularly the strongest ones (KF10). In this work we derived the depth-dependent \vmic\ profile at each observed pulsation phase.

The bottom panels of Fig.~\ref{fig:hydrogen_vmic} show the \vmic\ profile as a function of depth, around the main line forming region, at three specific points in the RR\,Lyr's pulsation cycle. The \vmic\ profile becomes steeper at the more turbulent phases showing the increasing impact of the pulsation on the physical conditions of the atmosphere.

\begin{figure*}
\includegraphics[width=17cm,clip]{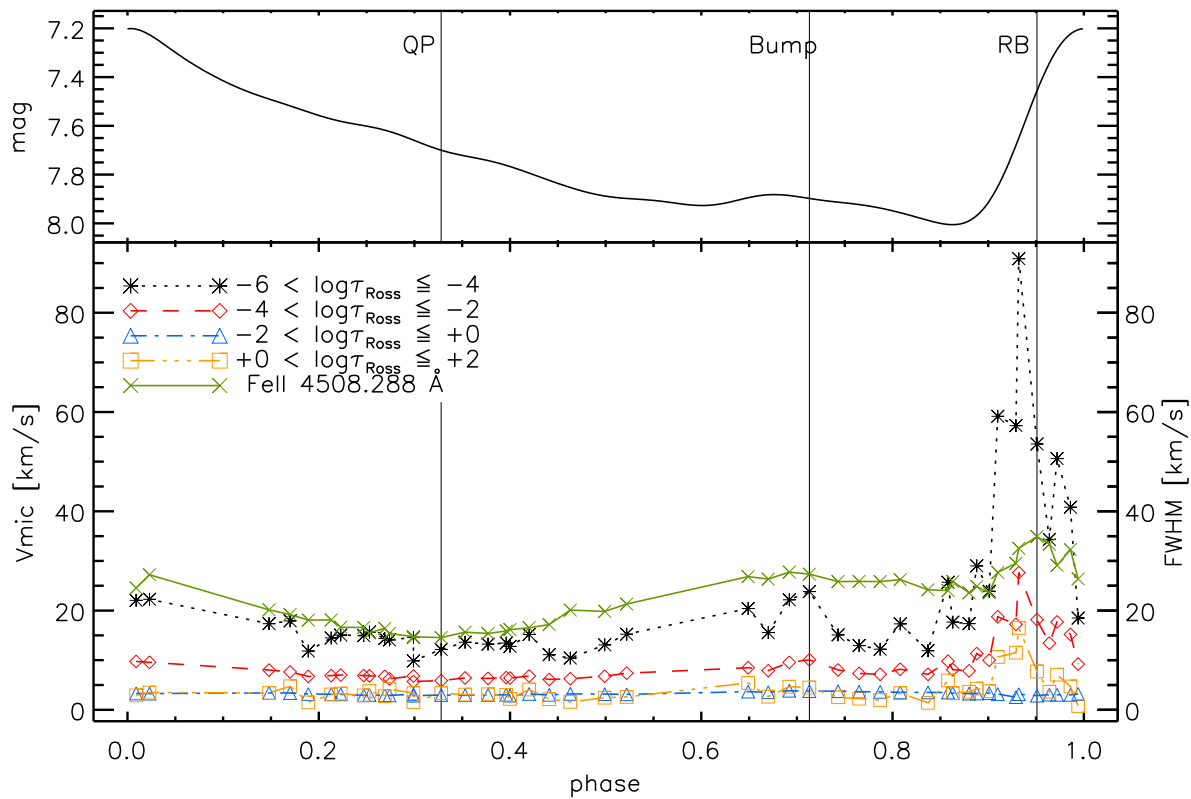}
\caption{Top panel: light curve (differential magnitude) fitted to the $V$-band photometry obtained simultaneously to the spectroscopic data (from K06). Bottom panel: average \vmic\ value in four different regions of the atmosphere. The black asterisks (connected by a dotted line), red rhombs (connected by a dashed line), blue triangles (connected by a dash-dotted line), and yellow squares (connected by a dash-triple-dotted line) show the average \vmic\ value at $-6<\log\tau_{Ross}\leq-4$, $-4<\log\tau_{Ross}\leq-2$, $-2<\log\tau_{Ross}\leq0$, $0<\log\tau_{Ross}\leq2$, respectively. The green crosses (connected by a solid line) show the full width at half maximum (FWHM) in \kms\ given by KF10 for the Fe\,{\sc ii} line at 4508.288\,\AA, which has its main line forming region at $-1.5\lesssim\log\tau_{Ross}\lesssim0.5$. Note that the \vmic\ profiles in the outermost layers might be an artifact of the polynomial extrapolation.}
\label{fig:vmic-fwhm}
\end{figure*}

Figure~\ref{fig:vmic-fwhm} allows one to follow the structure of the \vmic\ profile along the pulsation cycle. The plot shows the average \vmic\ value obtained in four different atmospheric regions as a function of pulsation phase. The \vmic\ profile close to the photospheric level ($-2<\log\tau_{Ross}\leq0$) is the least affected by the pulsation and remains almost constant during the pulsation cycle. This agrees with the fact that the adoption of a depth dependent \vmic\ becomes necessary when one considers the strongest lines, which are formed mostly higher up in the atmosphere. As a matter of fact, Fig.~\ref{fig:vmic-fwhm} suggests that in higher layers ($-6<\log\tau_{Ross}\leq-2$) \vmic\ steeply increases and its behaviour is clearly bound to the pulsation. We obtained the flattest \vmic\ profiles on the descending branch, while there is an increase in \vmic\ around the bump phase and again on the rising branch. In the inner-most layers, \vmic\ shows a phase dependence similar to that of the outer layers. Note that the \vmic\ profiles in the outermost and innermost layers might be an artifact of the polynomial extrapolation, as almost none of the measured metallic lines is formed at these depths.

\citet{fokin99} deduced the RR\,Lyr turbulent velocity as a function of phase from the observed variation of the FWHM of the Fe\,{\sc ii} line at $\sim$4923\,\AA. For a better comparison with \citet{fokin99}'s results we show in Fig.~\ref{fig:vmic-fwhm} also the FWHM of the Fe\,{\sc ii} line at 4508.288\,\AA, which was measured by KF10. Although the two lines are different, the comparison is justified by the fact that the two transitions belong to the same ion and have very similar excitation energies and strengths, hence a similar line formation depth. The general behaviour of the observed \vmic\ as a function of pulsation phase resembles that obtained by \citet{fokin99}, which appears to be relatively constant along the descending branch, with maxima at the bump phase and at maximum light. In addition, within the $-4\,\lesssim\,\log\tau_{Ross}\,\lesssim\,0$ atmospheric region, the range spanned by the average \vmic\ value is very similar to that given by \citet{fokin99}.

The major difference between the modelled and observed turbulent velocity variation is around the bump phase, where \citet{fokin99} obtained a very strong peak of the turbulent velocity, followed by a further smaller sharp peak on the rising branch. Observationally, we also obtained an increase in the turbulence velocity around the bump phase, but it is considerably smaller than the modelled one, and in particular it is smaller than the one on the rising branch. Since the FWHM of the two Fe\,{\sc ii} lines shown by \citet{fokin99} and here are very similar, the differences between the modelled and observed turbulence variation as a function of phase suggests that other parameters, beyond \vmic, play a role in shaping the spectral line profile variations along the pulsation cycle. In particular, the line asymmetries, which affect the FWHM, are not caused by small scale turbulent motions (\vmic) and, if not taken into account, might lead to a mis-interpretation of the FWHM in terms of turbulent velocity.

From a hydrodynamical point of view, the phases around 0.9 (middle of the rising branch) where both our results and hydrodynamic modelling \citep{fokin99} show a decrease of the turbulence velocity, deserve a further detailed observational and theoretical study which goes beyond the scope of this work. Although an analysis of the uncertainties on the derived \vmic\ profiles is still on-going and will be part of a separate work (Fossati et al., in preparation), we can already say that this decrease in \vmic\ is probably statistically significant. 

It is important to notice that for radial pulsators, such as RR~Lyr, also the radial velocity (\vr) is depth-dependent and a priori it is not easily possible to disentangle between a depth-dependent \vmic\ and a depth-dependent \vr; this is because a depth dependent \vr\ leads to variations in the line equivalent widths used to determine \vmic. One hint, however, would come from the shape of the line profiles, as only a depth-dependent \vr\ would produce asymmetric line profiles. We believe that our data show that RR~Lyr presents both a depth-dependent \vmic\ and a depth-dependent \vr. This is supported by the fact that the line profiles of RR~Lyr are asymmetric at most phases, but close to the ``quiet phase'', where the lines are most symmetric and therefore there should not be a strong depth-dependent \vr\ component, we indeed clearly see the signature of a depth-dependent \vmic. A further complication comes from the fact that in a RR~Lyrae star one cannot a priori exclude also the presence of a depth-dependent macroturbulence velocity \vmac. On the other hand, there should not be degeneracy between \vmac\ and \vmic, because in principle \vmac\ variations do not modify equivalent widths, that are instead used to measure \vmic. 

\citet{preston2013} derived the \vmic\ and \vmac\ values as a function of pulsation phase for the sample of stars analysed by \citet{for11}. They did not take into account any depth dependent component, hence their values can be considered as averages across the line-forming region. They obtained that \vmic\ and \vmac\ vary together with phase, where the difference is only in the scale of the variation. This further suggests that indeed the variation of both \vmic\ and \vmac\ has the same pulsational origin.

It is clear from the analysis presented in KF10 and here that the determination of the depth dependent \vmic\ profile is important to characterise the atmospheric structure and to more directly compare the observational results with hydrodynamic modelling. If one is constrained to use a constant \vmic\ value, we suggest to derive it from spectra obtained on the descending branch and to avoid using the strongest lines, most affected by \vmic\ variations. As a matter of fact \citet{fossati11} showed that, particularly for Fe\,{\sc i}, there is a strong relation between line strength and excitation energy, linking therefore the determination of \Teff\ and \vmic, when obtained from the equilibria of Fe\,{\sc i} lines. In agreement with this, \citet{takeda06} adopted a constant \vmic, but derived it avoiding the strongest Fe\,{\sc i} lines, hence they obtained almost the same \vmic\ value at all phases, except for the phase on the rising branch where they obtained a smaller \vmic\ value. This smaller value might be the result of adopting a too large \Teff\  (see Fig.~\ref{fig:LTR}) or because the spectrum was obtained at the phase (rising light) where we also derived a general decreasing \vmic. \citet{for11} derived \vmic\ for a number of RR\,Lyrae pulsators from the equilibrium of both Fe\,{\sc i} and Fe\,{\sc ii} lines, including strong lines. They adopted a constant \vmic\ value, though their line abundance vs. equivalent width equilibrium shows indeed the need of a depth dependent \vmic. As a consequence they obtained a \vmic\ value varying with phase on average as much as $\sim$1.5\,\kms\ and with a shape which plausibly resembles the picture we detailed in Fig.~\ref{fig:vmic-fwhm}, where the minimum \vmic\ values are obtained on the descending branch.
\section{Turbulence as origin for the Blazhko cycle?}
The modulation of the light curves of a large fraction of the RR~Lyrae stars (the Blazhko effect) still eludes a definitive and complete explanation. In the course of the past decade, both magnetic and resonance models involving non-radial modes were challenged by the observations. Following the results from the {\it Kepler} mission on RR~Lyrae stars, the radial resonance model proposed by \citet{bk} receives the widest acceptance, though the predictions of the resonance with the strange mode for the Blazhko effect remain to be verified with hydrodynamic models. A quoted alternative to the resonance model, and also of increasing popularity in recent years, is the scenario proposed by \citet{sto10}, in which variable turbulent convection (possibly as a consequence of a decaying magnetic field) leads to a variable quenching of the driving of the pulsation. This scenario has been quoted as providing an explanation for the quasi-periodic nature of the Blazhko effect as observed in several stars. More recently, \citet{Gillet13} proposed a model for the Blazhko effect in which a specific shock wave, called first overtone-shock, is generated by the perturbation of the fundamental mode by the transient first overtone. This shock causes a slowdown of the in-falling atmospheric layers and affects the intensity of the $\kappa$-mechanism. After an amplification phase, the intensity of the main shock reaches its highest critical value at the Blazhko maximum. The motion of the photospheric layers is desynchronised, after which the atmosphere relaxes and reaches a new synchronous state at Blazhko minimum. As \citet{Gillet13} writes, because of the non-linearity of the involved physical mechanisms (shocks, atmospheric dynamics, radiative losses, mode excitations), the Blazhko process is expected to be somewhat unstable and irregular, as we observe in several Blazhko stars.

Both the \citet{sto06,sto10} scenario and the shock model by \citet{Gillet13} have a direct connection with the turbulence in the star. Therefore, the microturbulent velocity \vmic\ as is derived in this paper, can be related to these hypotheses. Although it may only capture a fraction of the turbulence the star is actually undergoing, \vmic\ would reflect the extent of turbulent motions assumed in both the \citet{sto06,sto10} scenario and the \citet{Gillet13} shock model.

As described above, variations in the convective motion could be the origin of the Blazhko effect. In order to test this hypothesis \citep{sto06,sto10} to some extent, we used the derived depth dependent \vmic\ profiles to estimate the effect of the variable \vmic\ on the stellar magnitude and therefore on the light curve. We calculated synthetic fluxes assuming the atmospheric parameters we derived at the QP and adopting the \vmic\ profile obtained at the QP and another, very different one, on the rising branch (i.e., pulsation phase 0.951 / spectrum number 174). We then derived the Johnson $V$-band magnitude from the synthetic fluxes. In this way we measured the effect of a strong variation of the microturbulence profile in terms of stellar magnitude. The difference in $V$-band magnitude from the two synthetic fluxes is of $\sim$0.01\,mag, much smaller than typical Blazhko variations. This shows that even strong modifications in the microturbulent motions would not be able to explain the large magnitude variations shown by the Blazhko effect. We derived the magnitudes also in other photometric bands, obtaining that the difference due to the use of the two \vmic\ profiles is wavelength dependent and gradually decreases with decreasing wavelength: in the Johnson $B$ and $U$ bands that variation is of 0.007\,mag and 0.005\,mag, respectively. In all bands, the star with the ``quiet'' \vmic\ profile is fainter. 

As mentioned above, it remains to be investigated and quantified how much the microturbulence profiles capture the ``turbulence'' addressed in the models by \citet{sto06,sto10} and \citet{Gillet13}. Nevertheless, our test shows that much larger, and probably unrealistic, \vmic\ variations are needed to cause the observed Blazhko variations. This result is in line with, and complementary to, the findings by \citet{smolec2011}, who tested the effects of a variable turbulence in the form of the mixing-length parameter, on the light curves of RR~Lyrae stars, with similar conclusions regarding to the \citet{sto06,sto10} scenario.

On another note, the techniques adopted in this paper will allow us to investigate the model by \citet{Gillet13} in the needed detail. As \citet{Gillet13} points out, the intensity of the main shock wave before the Blazhko maximum becomes high enough to provoke large radiative losses, which can be at least equal to 70\% of the total energy flux of the shock. This would induce a small decrease of the effective temperature at each pulsation cycle. In a forthcoming study, we will use the methods detailed in this work to investigate this effect.
\section{Conclusions}
In this paper we present an in-depth spectroscopic analysis of RR~Lyr over a complete pulsation cycle. We build upon the findings of our analysis of the pulsation phase at maximum radius (the most ``quiet phase'', QP) using about the same methodology as described in \citet{KF10}. We present the results for a series of 49 spectra along the pulsation cycle of RR~Lyr at Blazhko phase 0.30--0.39, i.e. between Blazhko maximum and Blazhko minimum.
\begin{itemize}
\item The effective temperature \Teff, as determined by fitting the observed wings of the H$\gamma$ line, varies between 5950\,K and 7125\,K with uncertainties between 50\,K and 150\,K, slightly increasing with temperature, because of the decreasing sensitivity of hydrogen lines to temperature variations with increasing \Teff. The effective temperature shows good agreement with previous temperature determinations, where differences can be explained by improvements in the stellar atmosphere models and the effect of the Blazhko modulation. Even larger temperature variations can be expected at the phase of Blazhko maximum.

\item Determining the effective gravity \logg\ is not straightforward, particularly for RR~Lyr, as a consequence of line profile variations due to pulsation and because of non-LTE effects. From the non-detection of variations in the ionisation equilibrium along the pulsation cycle we estimate that \logg\ variations are small, within 0.1\,dex, hence we fixed \logg\ at 2.4, the value obtained at the QP \citep{KF10}.

\item In \citet{KF10} we derived a depth-dependent \vmic\ profile that allowed for a much better fit of all lines, especially for the strongest ones. In this work we derived the depth-dependent \vmic\ profile over the pulsation cycle, adopting a more representative set of lines compared to what given in \citet{KF10}. Due to the range of line formation depths covered by the measured lines, we estimate the \vmic\ profiles to be reliable in the optical depth range $-4\,\lesssim\,\log\tau_{Ross}\,\lesssim\,0$, which is the main line forming region. A thorough investigation of the reliability of the determination of the depth-dependent \vmic\ profile will be the topic of a following work (Fossati et al., in preparation). 

\item Element abundances resulting from our analysis are stable over the pulsation cycle, as one would expect. In that respect, our results are in good agreement with those of \citet{for11}. In addition, we can conclude that element abundances can be most safely determined from spectra taken on the descending branch of RR~Lyrae stars, even when not observed exactly at the phase when the spectra are least distorted (i.e., QP). 

\item  We observed that the abundance of singly ionised elements decreases at minimum light, reaches a minimum in the middle of the rising branch (around minimum radius) and goes back to the average value at maximum light. We attribute this to a variation of the dynamical acceleration term in the effective gravity.

\item Finally, assuming turbulence is at the origin of the Blazhko effect \citep{sto06,sto10}, we checked the effect of a variable \vmic\ profile on the stellar magnitude. We found that even large variations in the microturbulent motions (and hence \vmic\ profile) cannot explain the extent of the magnitude variations seen in Blazhko stars.  
\end{itemize}

\section*{Acknowledgments}
The authors would like to thank the referee George Preston for useful comments that improved the manuscript. LF acknowledges financial support from the Alexander von Humboldt Foundation. KK is supported by a Marie Curie International Outgoing Fellowship (255267 SAS-RRL) with the $7^{\rm th}$ European Community Framework Program. Part of this investigation has been supported by the Austrian Fonds zur F\"orderung der wissenschaftlichen Forschung, project number P17097-N02. DS acknowledges financial support from CRC~963~--~Astrophysical Flow Instabilities and Turbulence (project A16 and A17). AE acknowledges support from The Scientific and Technological Research Council of Turkey  (T\"{U}B\.{I}TAK), project number 112T119. OK is a Royal Swedish Academy of Sciences Research Fellow supported by grants from the Knut and Alice Wallenberg Foundation, the Swedish Research Council and G\"oran Gustafsson Foundation.
\appendix
\section{Phase-by-phase element abundances}
Table~\ref{tab:abn} lists the element abundances, in $\log(N_{X}/N_{tot})$, derived from each analysed spectrum.
\begin{landscape}
\begin{table*}
\caption[ ]{Element abundances, in $\log(N_{X}/N_{tot})$, derived from each analysed spectrum. The estimated internal errors in units of 0.01\,dex and the number of lines measured for each element are given in parenthesis. An abundance value of 0.00 was assigned to the elements for which no lines were measured.}
\label{tab:abn}
\begin{center}
\scriptsize{
\begin{tabular}{lccccccccccccc}
\hline
\hline
Pulsation & Sp. ID &Mg\,{\sc i}&Si\,{\sc i}&Si\,{\sc ii}&Ca\,{\sc i}&Ti\,{\sc i}&Ti\,{\sc ii}&Cr\,{\sc i}&Cr\,{\sc ii}&Fe\,{\sc i}&Fe\,{\sc ii}&Ni\,{\sc i}&Ba\,{\sc ii}\\ 
phase     & number &   [dex]   &   [dex]   &   [dex]	&   [dex]   &	[dex]	&   [dex]    &   [dex]   &   [dex]    &   [dex]   &    [dex]   &   [dex]   &   [dex]	\\
\hline
0.098 & 251 & -5.52(06/4) & -5.64(00/1) & -5.47(00/1) & -6.95(05/12) & -8.27(05/2) & -8.26(07/23) & -7.92(07/7) & -7.67(11/8) & -6.03(11/99) & -5.97(10/23) & -7.38(02/2)  &-11.51(06/4)\\
0.120 & 252 & -5.50(12/7) & -5.60(15/5) & -5.52(03/2) & -6.98(06/13) & -8.33(06/5) & -8.28(12/30) & -7.97(06/6) & -7.77(06/5) & -6.04(11/119)& -5.98(10/26) & -7.36(01/4)  &-11.49(04/4)\\
0.141 & 253 & -5.44(15/5) & -5.61(00/1) & -5.48(00/1) & -6.88(02/7)  & -8.29(00/1) & -8.25(08/14) & -7.80(11/3) & -7.64(14/3) & -6.05(10/78) & -5.93(08/14) & -7.32(02/2)  &-11.47(03/3)\\
0.173 & 087 & -5.52(09/5) & -5.67(12/3) & -5.53(07/3) & -6.96(04/15) & -8.27(02/5) & -8.22(08/31) & -7.95(05/7) & -7.64(10/11)& -6.02(10/130)& -5.94(08/27) & -7.32(05/8)  &-11.41(05/4)\\
0.173 & 255 & -5.51(06/5) & -5.55(09/4) & -5.46(00/1) & -6.94(06/14) & -8.30(05/5) & -8.22(06/16) & -7.93(09/5) & -7.65(13/5) & -6.01(12/112)& -5.98(08/16) & -7.31(07/8)  &-11.40(02/3)\\
0.203 & 256 & -5.52(08/5) & -5.57(13/9) & -5.54(04/3) & -6.92(07/16) & -8.28(07/11)& -8.28(12/27) & -7.92(06/8) & -7.72(01/11)& -6.03(11/145)& -5.94(08/26) & -7.32(09/14) &-11.36(08/4)\\
0.207 & 088 & -5.45(12/8) & -5.70(15/7) & -5.52(03/2) & -6.91(07/15) & -8.25(04/6) & -8.22(10/32) & -7.92(07/12)& -7.62(14/18)& -6.00(11/167)& -5.92(09/32) & -7.29(07/17) &-11.35(08/4)\\
0.226 & 257 & -5.53(10/6) & -5.60(10/5) & -5.58(00/1) & -6.91(06/14) & -8.31(01/4) & -8.29(11/20) & -7.89(09/5) & -7.65(10/10)& -6.03(11/120)& -5.94(11/22) & -7.34(08/8)  &-11.38(06/3)\\
0.229 & 089 & -5.52(11/8) & -5.57(13/10)& -5.58(14/3) & -6.91(07/16) & -8.24(10/7) & -8.22(08/25) & -7.92(05/12)& -7.55(09/15)& -6.00(11/153)& -5.89(08/28) & -7.29(07/14) &-11.33(08/4)\\
0.249 & 258 & -5.48(08/4) & -5.68(08/6) & -5.50(04/3) & -6.93(07/12) & -8.30(03/6) & -8.26(16/32) & -7.91(11/11)& -7.60(11/10)& -6.03(11/150)& -5.94(08/24) & -7.34(06/20) &-11.36(07/3)\\
0.260 & 091 & -5.56(11/8) & -5.63(12/6) & -5.54(10/3) & -6.94(06/16) & -8.30(03/5) & -8.29(12/34) & -7.90(08/9) & -7.69(16/14)& -6.01(11/175)& -5.92(08/28) & -7.27(07/20) &-11.37(02/4)\\
0.278 & 260 & -5.53(08/5) & -5.63(13/9) & -5.54(08/2) & -6.95(06/15) & -8.33(07/11)& -8.24(08/22) & -8.03(05/8) & -7.67(12/8) & -6.06(11/164)& -5.93(09/24) & -7.35(06/18) &-11.38(07/3)\\
0.303 & 261 & -5.47(06/4) & -5.73(11/7) & -5.54(07/3) & -6.94(08/16) & -8.28(08/12)& -8.29(12/31) & -7.99(10/9) & -7.66(11/8) & -6.06(10/135)& -5.93(08/23) & -7.35(09/18) &-11.38(06/3)\\
0.327 & 262 & -5.57(10/4) & -5.67(14/10)& -5.53(04/3) & -6.96(06/14) & -8.35(05/11)& -8.32(13/27) & -8.05(04/7) & -7.65(09/11)& -6.08(11/165)& -5.95(09/25) & -7.37(07/18) &-11.41(04/3)\\
0.349 & 204 & -5.49(11/5) & -5.66(15/8) & -5.54(12/2) & -6.94(08/12) & -8.30(02/8) & -8.25(08/20) & -7.95(05/8) & -7.66(10/7) & -6.05(11/143)& -5.94(08/25) & -7.32(05/13) &-11.34(06/3)\\
0.349 & 263 & -5.50(12/5) & -5.67(09/7) & -5.48(05/2) & -6.95(08/13) & -8.27(07/10)& -8.28(15/24) & -7.94(11/10)& -7.64(09/11)& -6.05(11/147)& -5.93(08/22) & -7.31(09/20) &-11.36(06/3)\\
0.372 & 205 & -5.55(11/5) & -5.56(07/6) & -5.61(07/2) & -6.96(06/14) & -8.35(06/6) & -8.30(09/29) & -7.95(08/9) & -7.70(12/6) & -6.06(12/146)& -5.96(10/24) & -7.33(05/13) &-11.36(10/4)\\
0.394 & 206 & -5.54(16/5) & -5.69(08/4) & -5.45(06/2) & -6.95(08/12) & -8.34(04/5) & -8.24(07/22) & -7.99(11/7) & -7.63(10/11)& -6.08(12/132)& -5.90(09/24) & -7.42(09/9)  &-11.42(01/2)\\
0.416 & 207 & -5.51(18/5) & -5.79(00/1) & -5.67(00/1) & -6.98(04/11) & -8.38(11/2) & -8.28(12/18) & -8.03(05/4) & -7.73(00/2) & -6.10(14/87) & -5.94(11/12) & -7.52(02/2)  &-11.41(04/3)\\
0.452 & 209 & -5.51(11/6) & -5.70(11/4) & -5.58(01/2) & -6.94(07/11) & -8.32(03/5) & -8.30(15/21) & -7.98(05/8) & -7.66(10/5) & -6.05(12/130)& -5.95(09/19) & -7.37(06/10) &-11.40(07/3)\\
0.475 & 210 & -5.51(16/5) & -5.51(03/3) & -5.53(08/2) & -6.94(06/13) & -8.33(08/8) & -8.24(08/22) & -7.92(07/9) & -7.64(12/8) & -6.05(12/146)& -5.95(08/19) & -7.37(09/7)  &-11.41(07/3)\\
0.604 & 158 & -5.57(12/7) & -5.76(01/2) & -5.62(10/2) & -7.04(07/10) & -8.36(00/1) & -8.35(09/18) & -8.02(08/5) & -7.79(14/5) & -6.14(10/86) & -6.07(10/19) & -7.53(00/1)  &-11.49(05/4)\\
0.626 & 159 & -5.47(14/7) & -5.53(00/1) & -5.53(01/2) & -6.99(09/9)  & -8.36(00/1) & -8.34(09/18) & -7.98(07/6) & -7.77(14/5) & -6.09(10/98) & -6.04(09/17) & -7.39(06/4)  &-11.46(06/4)\\
0.647 & 160 & -5.53(09/6) & -5.65(08/3) & -5.62(02/2) & -6.98(07/14) & -8.40(00/1) & -8.41(12/20) & -8.01(05/6) & -7.80(13/5) & -6.12(10/114)& -6.09(11/23) & -7.29(14/4)  &-11.53(04/3)\\
0.669 & 161 & -5.53(13/6) & -5.86(05/2) & -5.63(05/2) & -7.00(06/10) & -8.39(05/2) & -8.39(11/19) & -8.00(04/6) & -7.77(11/7) & -6.11(11/96) & -6.09(09/18) & -7.40(02/4)  &-11.52(03/3)\\
0.698 & 163 & -5.53(11/6) & -5.85(00/1) & -5.54(04/2) & -6.99(04/11) & -8.34(03/3) & -8.37(12/22) & -8.07(05/10)& -7.79(09/6) & -6.12(11/111)& -6.06(10/18) & -7.40(07/6)  &-11.54(04/3)\\
0.720 & 164 & -5.50(12/6) & -5.78(06/2) & -5.59(05/2) & -6.99(05/11) & -8.39(08/3) & -8.35(10/17) & -8.07(06/8) & -7.80(08/6) & -6.10(10/109)& -6.05(09/20) & -7.39(07/9)  &-11.49(09/3)\\
0.741 & 165 & -5.52(08/6) & -5.71(00/1) & -5.56(06/2) & -7.00(08/13) & -8.41(09/2) & -8.33(10/15) & -8.02(03/7) & -7.79(13/6) & -6.11(11/106)& -6.03(09/24) & -7.49(12/6)  &-11.49(08/3)\\
0.763 & 166 & -5.52(12/6) & -5.74(15/3) & -5.59(05/2) & -7.01(07/11) & -8.35(00/2) & -8.36(09/21) & -8.06(06/6) & -7.76(11/8) & -6.13(10/99) & -6.05(09/23) & -7.42(09/7)  &-11.47(07/4)\\
0.792 & 168 & -5.52(14/6) & -5.79(07/3) & -5.60(06/2) & -6.99(07/10) & -8.37(06/4) & -8.35(09/19) & -8.05(02/10)& -7.82(09/4) & -6.11(11/107)& -6.06(09/18) & -7.43(09/5)  &-11.51(06/3)\\
0.814 & 169 & -5.55(12/6) & -5.86(05/2) &  0.00(00/0) & -7.02(06/10) & -8.44(04/3) & -8.41(09/17) & -8.07(05/8) & -7.86(09/4) & -6.14(12/108)& -6.11(11/18) & -7.48(05/8)  &-11.54(05/3)\\
0.820 & 119 & -5.52(11/6) & -5.77(02/3) & -5.66(10/2) & -7.02(08/10) & -8.43(06/3) & -8.40(09/18) & -8.09(08/9) & -7.84(11/9) & -6.12(12/127)& -6.07(10/21) & -7.41(10/11) &-11.54(06/3)\\
0.834 & 170 & -5.54(10/6) & -5.79(03/3) & -5.59(05/3) & -7.04(07/10) & -8.45(07/4) & -8.41(14/18) & -8.10(06/8) & -7.84(11/6) & -6.17(11/109)& -6.11(10/19) & -7.43(10/8)  &-11.55(09/4)\\
0.846 & 120 & -5.58(15/5) & -5.85(04/4) & -5.70(00/1) & -7.04(06/12) & -8.46(02/4) & -8.49(08/17) & -8.09(04/6) & -7.89(10/9) & -6.15(13/108)& -6.19(10/21) & -7.45(08/8)  &-11.67(01/2)\\
0.856 & 171 & -5.55(08/6) & -5.76(02/2) & -5.68(07/2) & -7.05(07/10) & -8.47(07/3) & -8.51(09/15) & -8.09(03/6) & -7.90(10/8) & -6.16(12/95) & -6.20(10/19) & -7.39(08/6)  &-11.69(02/4)\\
0.868 & 121 & -5.57(12/6) & -5.93(07/3) & -5.72(07/2) & -7.02(04/9)  & -8.39(00/1) & -8.55(07/18) & -8.10(06/7) & -7.92(10/9) & -6.14(12/97) & -6.27(09/21) & -7.51(06/5)  &-11.75(01/2)\\
0.885 & 173 & -5.49(11/5) & -5.65(00/1) & -5.68(00/1) & -7.02(09/12) &  0.00(00/0) & -8.50(13/12) & -8.01(10/2) & -7.85(08/5) & -6.14(10/71) & -6.22(06/15) & -7.48(03/2)  &-11.74(15/3)\\
0.890 & 122 & -5.55(16/6) & -5.25(00/1) & -5.75(02/2) & -7.00(04/10) & -8.39(00/1) & -8.59(09/14) & -7.98(14/3) & -8.04(14/5) & -6.09(11/63) & -6.32(11/21) & -7.44(02/2)  &-11.75(12/3)\\
0.905 & 174 & -5.39(06/4) &  0.00(00/0) & -5.54(09/2) & -6.98(07/10) &  0.00(00/0) & -8.53(07/11) & -8.03(01/2) & -7.80(03/6) & -6.08(11/55) & -6.16(08/23) & -7.41(09/2)  &-11.74(08/3)\\
0.922 & 124 & -5.52(10/4) &  0.00(00/0) & -5.64(09/3) & -6.99(04/7)  &  0.00(00/0) & -8.54(10/15) & -7.99(08/2) & -7.98(11/5) & -6.08(14/49) & -6.23(11/23) & -7.43(00/1)  &-11.75(15/3)\\
0.928 & 175 & -5.52(14/5) & -5.49(00/1) & -5.49(00/1) & -6.95(07/8)  &  0.00(00/0) & -8.51(07/12) & -8.01(08/2) & -7.87(12/7) & -6.07(09/52) & -6.19(11/20) &  0.00(00/0)  &-11.74(14/3)\\
0.943 & 125 & -5.46(12/5) & -5.50(00/1) & -5.53(04/2) & -6.95(07/6)  &  0.00(00/0) & -8.42(09/20) & -7.83(01/2) & -7.85(07/7) & -6.03(15/58) & -6.15(09/19) &  0.00(00/0)  &-11.69(14/3)\\
0.948 & 176 & -5.50(09/6) &  0.00(00/0) & -5.46(00/1) & -6.98(09/13) & -8.34(00/1) & -8.43(05/14) & -8.04(10/2) & -7.80(12/9) & -6.06(09/67) & -6.12(09/24) & -7.31(11/2)  &-11.73(11/3)\\
0.967 & 126 & -5.48(12/6) &  0.00(00/0) & -5.52(01/2) & -6.96(09/5)  &  0.00(00/0) & -8.35(09/16) & -8.01(00/1) & -7.94(12/2) & -6.04(10/26) & -6.00(12/21) &  0.00(00/0)  &-11.55(26/2)\\
0.977 & 178 & -5.50(04/5) &  0.00(00/0) &  0.00(00/0) & -7.01(06/9)  & -8.29(00/1) & -8.35(05/15) & -7.92(15/3) & -7.76(11/8) & -6.05(09/76) & -6.06(10/24) & -7.34(00/1)  &-11.69(11/3)\\
\hline
\end{tabular}}
\end{center}
\end{table*}
\end{landscape}


\bsp

\label{lastpage}

\end{document}